\documentclass[letterpaper,twocolumn,prl,aps,superscriptaddress,amsmath,amssymb,floatfix]{revtex4-1}
\usepackage{mathptmx}
\usepackage[latin9]{inputenc}
\usepackage{color}
\usepackage{verbatim}
\usepackage{float}
\usepackage{amsmath}
\usepackage{amssymb}
\usepackage{graphicx}
\usepackage{esint}
\usepackage[T1]{fontenc}
\usepackage[unicode=true,
 bookmarks=true,bookmarksnumbered=false,bookmarksopen=false,
 breaklinks=false,pdfborder={0 0 1},backref=false,colorlinks=true]
 {hyperref}
\hypersetup{
 linkcolor=magenta,urlcolor=blue,citecolor=blue,pdfstartview={FitH},hyperfootnotes=false}

\makeatletter

\usepackage{textcomp}
\usepackage{epstopdf}

\pdfpageheight\paperheight
\pdfpagewidth\paperwidth

\providecommand{\tabularnewline}{\\}

\@ifundefined{textcolor}{}{%
 \definecolor{BLACK}{gray}{0}
 \definecolor{WHITE}{gray}{1}
 \definecolor{RED}{rgb}{1,0,0}
 \definecolor{GREEN}{rgb}{0,1,0}
 \definecolor{BLUE}{rgb}{0,0,1}
 \definecolor{CYAN}{cmyk}{1,0,0,0}
 \definecolor{MAGENTA}{cmyk}{0,1,0,0}
 \definecolor{YELLOW}{cmyk}{0,0,1,0}
}

\usepackage{xcolor}
\usepackage{soul}
\definecolor{blue}{rgb}{0,0,1}
\definecolor{red}{rgb}{1,0,0}
\definecolor{green}{rgb}{0,1,0}

\makeatother

\begin{document}

\affiliation{Institute of quantum information and technology, Nanjing University of Posts and Telecommunications, Nanjing 210003, China}
\affiliation{CAS Key Laboratory of Quantum Information, University of Science and Technology of China, Hefei 230026, China}
\affiliation{State Key Laboratory of Quantum Optics and Quantum Optics Devices, and Institute of Opto-Electronics, Shanxi University, Taiyuan 030006,
China}
\affiliation{Broadband wireless communication and sensor network technology, key lab of Ministry of Education, Nanjing University of Posts and Telecommunications, Nanjing 210003, China}
\affiliation{CAS Center for Excellence in Quantum Information and Quantum Physics, University of Science and Technology of China, Hefei 230026, China}

\title{Proposal for stable atom trapping on a GaN-on-Sapphire chip}
\author{Aiping Liu}
\affiliation{Institute of quantum information and technology, Nanjing University of Posts and Telecommunications, Nanjing 210003, China}
\affiliation{Broadband wireless communication and sensor network technology, key lab of Ministry of Education, Nanjing University of Posts and Telecommunications, Nanjing 210003, China}
\author{Lei Xu}
\affiliation{CAS Key Laboratory of Quantum Information, University of Science and
Technology of China, Hefei 230026, China}
\affiliation{CAS Center for Excellence in Quantum Information and Quantum Physics,
University of Science and Technology of China, Hefei 230026, China}
\author{Xin-Biao Xu}
\affiliation{CAS Key Laboratory of Quantum Information, University of Science and
Technology of China, Hefei 230026, China}
\affiliation{CAS Center for Excellence in Quantum Information and Quantum Physics,
University of Science and Technology of China, Hefei 230026, China}

\author{Guang-Jie Chen}
\affiliation{CAS Key Laboratory of Quantum Information, University of Science and
Technology of China, Hefei 230026, China}
\affiliation{CAS Center for Excellence in Quantum Information and Quantum Physics,
University of Science and Technology of China, Hefei 230026, China}

\author{Pengfei Zhang}
\affiliation{State Key Laboratory of Quantum Optics and Quantum Optics Devices,
and Institute of Opto-Electronics, Shanxi University, Taiyuan 030006,
China}

\author{Guo-Yong Xiang}
\affiliation{CAS Key Laboratory of Quantum Information, University of Science and
Technology of China, Hefei 230026, China}
\affiliation{CAS Center for Excellence in Quantum Information and Quantum Physics,
University of Science and Technology of China, Hefei 230026, China}
\author{Guang-Can Guo}
\affiliation{CAS Key Laboratory of Quantum Information, University of Science and
Technology of China, Hefei 230026, China}
\affiliation{CAS Center for Excellence in Quantum Information and Quantum Physics,
University of Science and Technology of China, Hefei 230026, China}

\author{Qin Wang}
\email{qinw@njupt.edu.cn}
\affiliation{Institute of quantum information and technology, Nanjing University
of Posts and Telecommunications, Nanjing 210003, China}
\affiliation{Broadband wireless communication and sensor network technology, key
lab of Ministry of Education, Nanjing University of Posts and Telecommunications,
Nanjing 210003, China}

\author{Chang-Ling Zou}
\email{clzou321@ustc.edu.cn}
\affiliation{CAS Key Laboratory of Quantum Information, University of Science and
Technology of China, Hefei 230026, China}
\affiliation{CAS Center for Excellence in Quantum Information and Quantum Physics,
University of Science and Technology of China, Hefei 230026, China}
\affiliation{State Key Laboratory of Quantum Optics and Quantum Optics Devices,
and Institute of Opto-Electronics, Shanxi University, Taiyuan 030006,
China}

\date{\today}

\begin{abstract}
The hybrid photon-atom integrated circuits, which include photonic microcavities and trapped single neutral atom in their evanescent field, are of great potential for quantum information processing. In this platform, the atoms provide the single-photon nonlinearity and long-lived memory, which are complementary to the excellent passive photonics devices in conventional quantum photonic circuits. In this work, we propose a stable platform for realizing the hybrid photon-atom circuits based on an unsuspended photonic chip. By introducing high-order modes in the microring, a feasible evanescent-field trap potential well $\sim0.3\,\mathrm{mK}$ could be obtained by only $10\,\mathrm{mW}$-level power in the cavity, compared with $100\,\mathrm{mW}$-level power required in the scheme based on fundamental modes. Based on our scheme, stable single atom trapping with relatively low laser power is feasible for future studies on high-fidelity quantum gates, single-photon sources, as well as many-body quantum physics based on a controllable atom array in a microcavity.
\end{abstract}
\maketitle

\section{Introduction}

With high scalability and enhanced light-matter interactions, the integrated quantum photonic chip becomes a promising tendency for study of quantum optics and quantum information processing~\cite{Kim2020,Elshaari2020,Pelucchi2022}. In the past decades, great progress has been achieved in photonic integrated circuits (PIC), with most attention being paid to the excellent photonic material platforms~\cite{Dai2012,Bogaerts2020,Shastri2021,Siew2021}. Various quantum photonic devices with passive components and probabilistic state generation are demonstrated, with the single emitters are strongly desired for single-photon level optical nonlinearity, which is the essential resource for quantum light sources and the deterministic quantum operations~\cite{Kimble08}. Neutral atoms with long-lived energy levels provide a potential solution for the essential single-level nonlinearity to the PIC, which could also provide the long-coherence time memory for storing quantum information. Besides, quantum computing and quantum sensing based on atoms can also be performed with the help of nanophotonic devices~\cite{chang2018colloquium,GarridoAlzar2019}.

Since there is a promising prospect for the photon-atom interaction on a PIC, considerable efforts have been recently devoted to combining the nanophotonic devices and neutral atoms to realize efficient quantum devices. Although much exciting progress has been achieved by the probabilistic strong coupling between the optical microcavities~\cite{Dayan2008,Shomroni2014,Volz2014}, it is important to realize stable quantum photonics devices to trap atoms on the nanophotonic devices. Various approaches have been developed to use near-field optical dipole traps to confine atoms at the surface of waveguide structures, including the nanofiber~\cite{6Corzo,Daly14,Meng2020},  optical waveguide~\cite{Goban2014,Luan2020,3Beguin,Gehl21}, the photonic crystal nanocavities~\cite{Tiecke2014,Samutpraphoot2020,Dorevic2021}, and also the microsphere resonator~\cite{Will2021}. However, these structures are suspended in a vacuum, so they are  potentially sensitive to vibrations and vulnerable to thermal instability. Alternatively, the atom trapping by unsuspended waveguide and microring structures have been proposed~\cite{Kohnen2010,Meng2015,Chang2019} and demonstrated~\cite{Kim2019,chang2020efficiently,Zhou2021}. These works of trapping atoms promote the realization of a stable and scalable hybrid PIC on a substrate for future applications.

In this work, a stable platform for realizing the hybrid photon-atom circuit based on a suspension-free gallium nitride (GaN) chip is proposed. The sapphire substrate is transparent to visible lasers, thus allowing efficient optical access for laser cooling and optical convey belt of cold atoms. To trap atoms, high-order guided modes in the microring waveguide are used to obtain a stable atom trapping near the surface of the microring. The required power of trapping light by hybrid modes trapping is reduced compared to the fundamental modes trapping. The manipulation of the atom states with waveguide mode is analyzed numerically, which indicates a strong interaction between the trapped atom and the confined optical photons through the evanescent field of waveguide mode. Such a GaN-on-Sapphire platform is feasible for hybrid photon-atom circuits and is promising for various quantum photonic devices.

\begin{figure*}
\includegraphics[width=1.6\columnwidth]{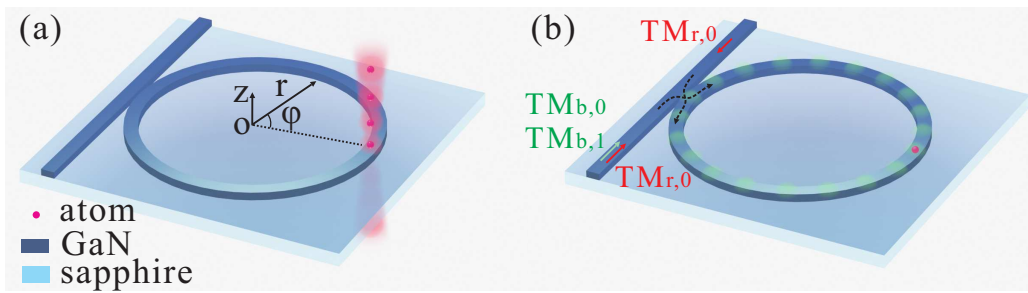}
\caption{Illustration of the microring platform for photon-atom interaction. (a) The microring platform with the atoms loaded to the surface of microring by the free space laser beam. (b) The microring platform with the atoms trapped on the surface of microring by the optical evanescent fields formed by the guiding modes.}
\label{fig1}
\end{figure*}

\section{The platform for hybrid photon-atom circuit}

The proposed experimental platform for hybrid photon-atom circuits is illustrated in Fig.~\ref{fig1}. The platform is made of GaN microrings and waveguides on a sapphire substrate, which is transparent to visible light. Such a platform has been extensively studied over the past decades, showing its excellent optical properties in visible wavelengths~\cite{Zhang2011,Bruch2015,Chen2017,Li2018,Stassen19} and also its potential in photonic integrated circuits~\cite{Fu2019,Xu2022}. Distinct from the conventional platforms for photon-atom interaction on a silicon substrate, all the photonic microstructures in our system are fabricated on the substrate, without any suspended structure or any fabrication procession of the substrate. The hybrid photon-atom circuits are prepared in three steps:

(i) Place the GaN-on-sapphire chip in vacuum. Since all the materials are transparent to visible light, the conventional magneto-optical trap could be implemented by transmitting laser beams through the chip. A cloud of cold atoms could be prepared by Doppler cooling, with the atomic cloud being only a few $mm$ away from the top suface of the microring device.

(ii) Guide the cold atoms to the chip. By sending a pair of counter-propagating focused Gaussian beams of red-detuned laser vertically through the transparent substrate, with the waist of the laser beams close to the microring on the chip, the cold atoms could be trapped and transported to the top surface of the microring by the optical conveyor belt~\cite{schrader2001optical, kuhr2001deterministic}. As illustrated in Fig.\,\ref{fig1}(a), by tuning the phase of upward and downward dipole laser beams, atoms could be guided and transported to the chip.

(iii) When the cold atoms are guided to a certain distance from the top surface of the microring, turn on the blue- and red-detuned lasers in the photonic integrated circuits, thus creating the two-color optical dipole trap~\cite{LeKien2004,Vetsch2010,fu2008atom,Chang2019} via the evanescent fields of the microring resonator. Then, the single atoms could be confined to the microring resonator, as shown in Fig.\,\ref{fig1}(b), and coupled to the signal photons near the transition frequency of the atoms.

Following this procedure, our platform offers a basic cavity-QED system for realizing the atom-photon quantum gates, atom-mediated photon-photon entanglement, as well as quantum storage.  It is also worth noting that the materials of this platform hold several advantages: Firstly, the unsuspended photonic circuit platform is very stable and is robust to the external mechanical perturbations and the thermal instability. Especially, to provide a deep trap for atoms by the evanescent field, a strong laser intensity in the microring should be excited, which will result in serious heating~\cite{padmaraju2012thermal,PadmarajuBergman14}. Secondly, since the GaN and sapphire are wide-bandgap (0.78-4.77 eV) materials transparent to a ultra-broad band wavelengths (260\,nm-1590\,nm) ~\cite{yu1997optical,muth1999absorption}, thus the scheme is valid for most atoms working with visible and near-visible lasers. Thirdly, for atomic transition at visible wavelengths, the refractive indices of the GaN and substrate are around $n_{1}=2.2$ and $n_{2}=1.7$, respectively. The large refractive index contrast indicates strong optical confinement in the GaN microphotonic structures~\cite{zheng2020integrated}. Lastly, GaN is not only promising for nonlinear photonics with large intrinsic nonlinearities and a large bandgap, but also quite resilient to high temperature and optical power. So the GaN microring can sustain a relative high intracavity optical power for the efficient manipulation of the neutral atoms.

As all the essential components and technique of the hybrid photon-atom circuit are feasible for experiments, here we are focusing on the design of the photonic structure for reasonable dipole traps and also for strong photon-atom interactions. So, in the following, we will investigate the performance of the dipole trap in detail and the potential photon-atom coupling strength.

\section{The simulation of the trap potential well}

\begin{figure}
\centering\includegraphics[width=1\columnwidth]{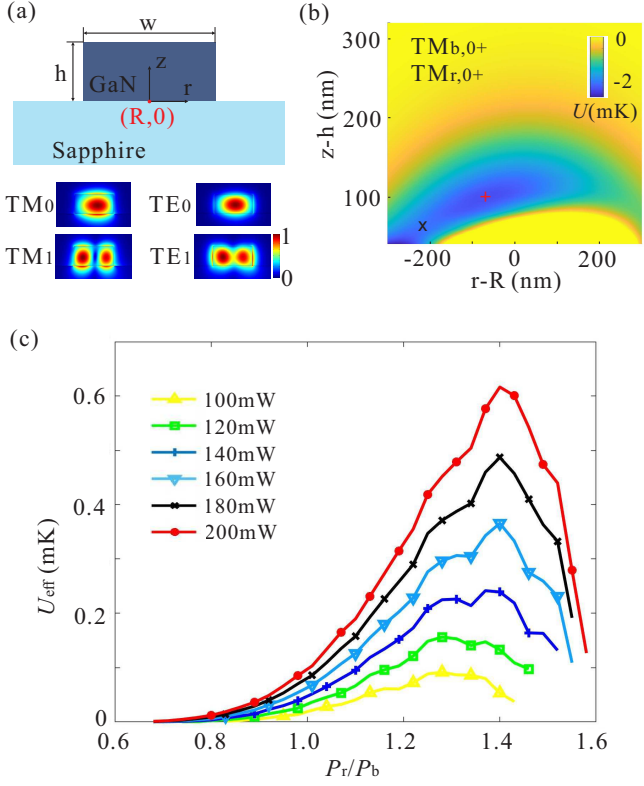}
\caption{The optical trap formed with the fundamental mode of microring.(a) The cross-section of the microring (up) and the optical field distributions of the propagating modes in the microring (down). (b) The optical potential distribution formed by the fundamental modes in the $r-z$ plane above the microring surface. (c) The optical traps vary with $P_\mathrm{r}/P_\mathrm{b}$ for different $P_\mathrm{b}$.}
\label{fig2}
\end{figure}

In this work, we consider the Rubidium atoms with the D2 transition wavelength of $^{87}$Rb atoms $\lambda_{0}=780\,\mathrm{nm}$. The optical properties of the GaN microring are determined by the geometry, which could be described by the major radius $R$, width $w$ and height $h$ of the cross-section of the microring, as shown in the up part of Fig.\,\ref{fig2}(a) , where $r=10\,\mathrm{\mu m},w=600\,\mathrm{nm}$, and $h=280\,\mathrm{nm}$. Two lasers, with the wavelengths $\lambda_{b}=760\,\mathrm{nm}$ and $\lambda_{r}=852\,\mathrm{nm}$, respectively, are serving as the blue- and red-detuned dipole trap light~~\cite{LeKien2004,Vetsch2010,fu2008atom,Chang2019}. When the two lasers are loaded into the microring, the excited microring modes could enhance the electric field and their evanescent parts in vacuum to offer the trap potential. The evanescent field decays exponentially away from the surface of the microring, so it provides a repulsive and a attractive force on the considered $^{87}$Rb atom by the blue- and red-detuned light modes, respectively. In addition, the attractive van der Waals force increases fast when the atom is close to the surface of the microring, thus the atoms will stick to the surface. Benefiting from the resonance enhancement of the microring, a strong evanescent field can be obtained by a relatively weak laser input to compensate for the van der Waals force. By appropriately manipulating the incident light, the dipole trap potential of the evanescent field and the van der Waals potential will form a stable potential well for trapping $^{87}$Rb atoms. For a balance between these optical forces, a trap position $100\,\mathrm{nm}$ away from the microring surface is considered in the following studies.

There are different waveguide modes in the microring waveguide, as illustrated in the inset of Fig.\,2(a) is the electric field distributions of the fundamental modes $\mathrm{TM}_0$ and $\mathrm{TE}_0$, as well as high-order modes $\mathrm{TM}_1$ and $\mathrm{TE}_1$ modes~\cite{XJ21} for incident light wavelength $\lambda_{b}$. Compared with $\mathrm{TE}$ modes, $\mathrm{TM}$ modes have a stronger evanescent field on the
top surface of waveguide, so only $\mathrm{TM}$ modes are considered in the following.
Shown in Fig.\,\ref{fig2}(b) is the cross section of simulated trap potential formed on top of the microring, in the simplest case that fundamental mode $\mathrm{TM}_0$ is excited for blue- and red-detuned lasers. Similar to the pioneer studies by Ref.~\cite{Chang2019}, the trap potential well is a narrow crescent because of the asymmetric field distributions of the microring modes on the microring cross section. The center of the trap potential well is marked by the red cross ``$+$" and the saddle point is marked by the black cross ``$\times$", respectively.
An effective trap potential of $0.265\,\mathrm{mK}$ (the potential difference between the well and saddle points) is obtained on the designed platform with $150\,\mathrm{mW}$ blue-detuned and $195\,\mathrm{mW}$ red-detuned $\mathrm{TM}_0$ laser power inside the microring. It is worth noting that the potential is evaluated for a given intracavity power, which should be much higher than the input power in the waveguide due to the resonant enhancement by the microring cavity modes. For example, with a practical quality factor of $Q\approx 2\times10^{6}$ and a free-specral range of $FSR\approx 320\,\mathrm{GHz}$~\cite{zheng2020integrated,Stassen:19}, the required input laser power to the microring could be only about $0.4\,\mathrm{mW}$.

The dependence of optical trap depth $U_\mathrm{eff}$ on the power of the circulating laser power inside the microring is shown in Fig.~\ref{fig2}(c), where $P_\mathrm{b}$ and $P_\mathrm{r}$ are the powers of the blue- and red-detuned $\mathrm{TM}_0$ modes circulating in the waveguide, respectively. The curves illustrate the optical trap depth $U_\mathrm{eff}$ varying with the power ratio of red- and blue-detuned $\mathrm{TM}_0$ modes ($P_\mathrm{r}/P_\mathrm{b}$), and the different curves correspond to the power of the blue-detuned $\mathrm{TM}_0$ mode as $100\,\mathrm{mW}$, $120\,\mathrm{mW}$, $140\,\mathrm{mW}$, $160\,\mathrm{mW}$, $180\,\mathrm{mW}$, and $200\,\mathrm{mW}$, respectively. From Fig.~\ref{fig2}(c), the optical trap depth is not only increased with the $P_\mathrm{b}$, but also related to the $P_\mathrm{r}/P_\mathrm{b}$. With a constant $P_\mathrm{b}$, the optical trap depth $U_\mathrm{eff}$ increases at first and then decreases when the $P_\mathrm{r}/P_\mathrm{b}$ increases. In fact, the optical trap can not be formed when the $P_\mathrm{r}/P_\mathrm{b}$ is too small or too large especially with low mode power. Besides of high enough intracaity powers, an appropriate $P_\mathrm{r}/P_\mathrm{b}$ is also required to form a large effective optical trap depth $U_\mathrm{eff}$.

\begin{figure}
\centering\includegraphics[width=1\columnwidth]{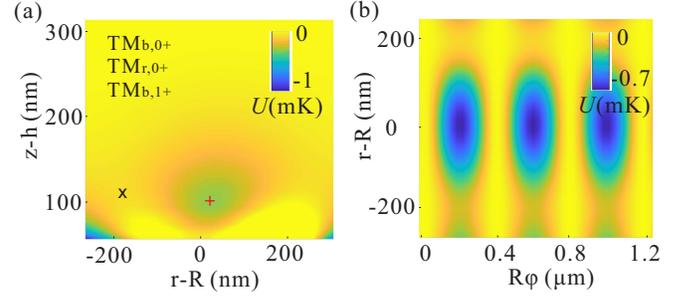}
\caption{The optical trap formed with hybrid modes. (a) The optical trap potential well above the microring surface in the $r-z$ plane. (b) The optical trap potential well formed with hybrid modes combined with bidirectional red-detuned $\mathrm{TM}_0$ modes and unidirectional blue-detuned $\mathrm{TM}_0$ and $\mathrm{TM}_1$ modes in the $r-\phi$ plane with $100\,\mathrm{nm}$ above the microring surface.}
\label{fig3}
\end{figure}

\begin{figure}
\centering\includegraphics[width=1\columnwidth]{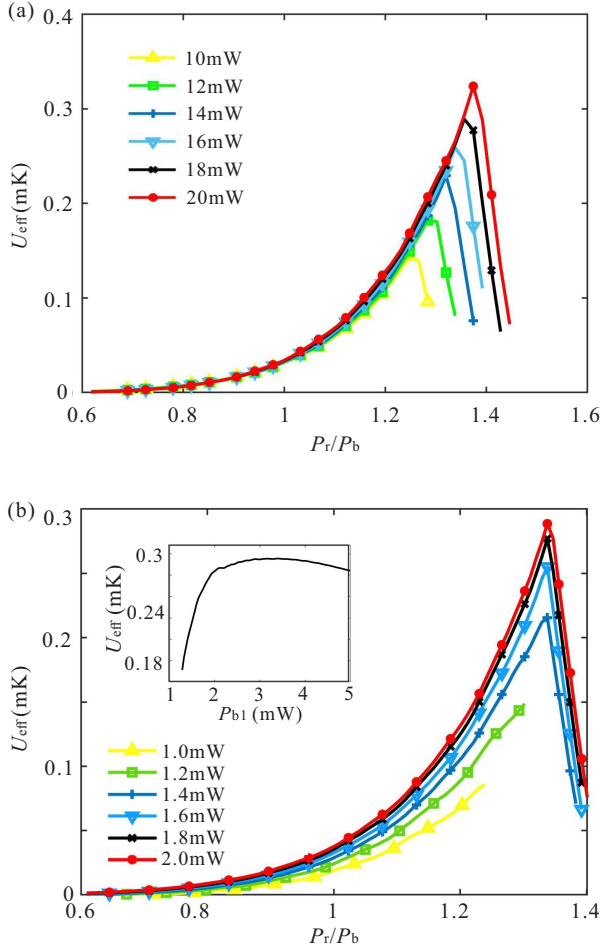}
\caption{ (a) The optical traps vary with $P_\mathrm{r}/P_\mathrm{b}$ for different $P_\mathrm{b}$ with $P_\mathrm{b1}=1.6\,\mathrm{mW}$.(b) The optical traps vary with $P_\mathrm{r}/P_\mathrm{b}$ for different $P_\mathrm{b1}$ with $P_\mathrm{b}=15\,\mathrm{mW}$. The inset gives the relation between the optical trap and $P_\mathrm{b1}$ with $P_\mathrm{b}=15\,\mathrm{mW}$ and $P_\mathrm{r}/P_\mathrm{b}=1.335$.}
\label{fig4}
\end{figure}

From the trap potential distribution in Fig.~\ref{fig2}(b), the effective trap depth is limited by the relative low trap depth at the saddle point. In order to further reduce the required laser power, we should improve the trap depth at the saddle point and obtain a more regular optical trap potential well. Therefore, a set of modes composed of blue-detuned $\mathrm{TM}_0$ mode, red-detuned $\mathrm{TM}_0$ mode and blue-detuned $\mathrm{TM}_1$ mode are used to form the optical trap potential well. As shown in Fig. \ref{fig3}(a) is the optical trap potential well formed with these hybrid modes, where the saddle point has a much higher potential barrier because the bivalve field of the assisted $\mathrm{TM}_1$ mode makes up the weak field on the two sides of the $\mathrm{TM}_0$ mode. The optical trap depth, with the trap center marked by the red cross ``$+$" and the saddle point marked by the black cross ``$\times$", is $0.272\,\mathrm{mK}$ with $P_\mathrm{b}=15\,\mathrm{mW}$, $P_\mathrm{r}/P_\mathrm{b}=1.335$, and the blue-detuned $\mathrm{TM}_1$ mode power $P_\mathrm{b1}=1.6\,\mathrm{mW}$. Further, the relations between the $U_\mathrm{eff}$ and the $P_\mathrm{r}/P_\mathrm{b}$ for different $P_\mathrm{b}$ are provided in Fig.~\ref{fig4}(a). The total mode power in the hybrid modes trapping scheme is reduced by one order of magnitude compared with the trapping scheme using the fundamental modes. With reduced incident light power, the potential thermal effect or mechanical vibration of the microring will be greatly suppressed, which makes the optical trapping more stable. Then a stable atom trap can be realized on the plane perpendicular to the microring waveguide.

The relation between the optical trap depth $U_{\mathrm{eff}}$ and the $P_\mathrm{r}/P_\mathrm{b}$ with $P_b=15\,\mathrm{mW}$ is given in Fig.~\ref{fig4}(b), where different curves correspond to different $P_\mathrm{b1}$. With increased $P_\mathrm{r}/P_\mathrm{b}$, the optical trap depth $U_\mathrm{eff}$ increases at first and then decreases after a maximun $U_\mathrm{eff}$. From different curves, the optical trap depth $U_\mathrm{eff}$ is larger with a bigger $P_\mathrm{b1}$, so is the maximun $U_\mathrm{eff}$. The inset of Fig.~\ref{fig4}(b) gives the relation between the optical trap depth and the blue-detuned $\mathrm{TM}_1$ mode power $P_\mathrm{b1}$ with $P_\mathrm{b}=15\,\mathrm{mW}$ and $P_\mathrm{r}/P_\mathrm{b}=1.335$. As the power of the blue-detuned $\mathrm{TM}_1$ mode $P_\mathrm{b1}$ increases, the optical trap depth increases fast at first to reach a maximum, and then decreases slowly. Thus, an appropriate $P_\mathrm{b}$ makes the optical trap depth more efficient and regular, thus stable and require low power consumption.

In addition, the bidirectional red-detuned $\mathrm{TM}_0$ modes are used to form the stationary field with other microring modes as unidirectional circulating waves. The optical trap potential well formed on top of the microring is given in Fig.~\ref{fig3}(b). It should be noted that, because of the material dispersion the wavevectors of the blue- and red-detuned circulating modes are different, only red-detuned $\mathrm{TM}_0$ mode is bidirectional circulating. Thus a stable atom trapping in three dimensions can be realized by the hybrid modes trapping on the platform based on suspension-free GaN-on-sapphire chip.

\section{Photon-atom interaction}

\begin{figure}
\centering \includegraphics[width=1\columnwidth]{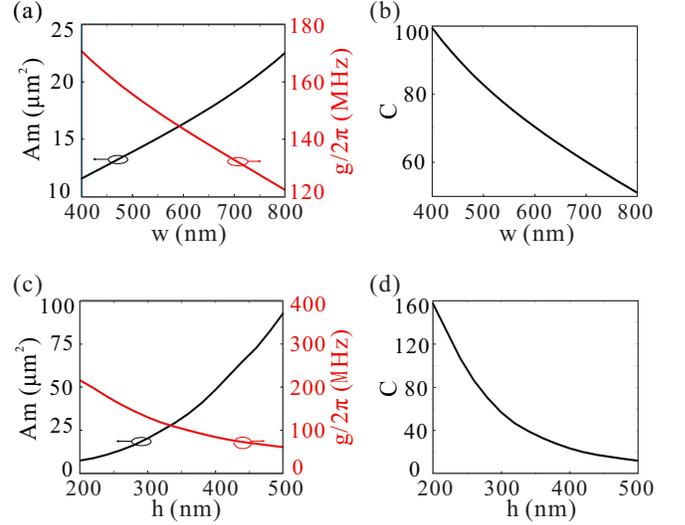}
\caption{The performance of microcavity enhanced atom-photon interaction of $^87$Rb in the trap center T($r=R$, $z-h=100\,\mathrm{nm}$). (a) The mode area $A_m$, and coupling strength $g$ (b)  cooperativity $C$ vary with the waveguide width $w$ of the microring with radius $R=10\,\mathrm{\mu m}$ and $h=280\,\mathrm{nm}$. (c) The mode area $A_m$, and coupling strength $g$ (d)  cooperativity $C$ vary with the waveguide height $w$ of the microring with $R=10\,\mathrm{\mu m}$ and $w=600\,\mathrm{nm}$. }
\label{fig5}
\end{figure}

The evanescent field around the microring waveguide will interact with the atoms trapped around the microring waveguide. Benefited from the low loss of GaN and excellent processing technology, the photonic loss rate ($\kappa$) can be smaller than the photon-atom coupling strength $g$, so is the atomic radiative decay rate ($\gamma$) into free space. Thus, the photon-atom strong  coupling regime is potentially achievable in current platform. For a perfect quantum emitter, the photon-atom coupling strength can be given as~\cite{Corwin99,Chang2019}
\begin{equation}
    g=\sqrt{\frac{3\lambda^{3}\omega_{0}\gamma}{16\pi^{2}V_{m}}},
\end{equation}
where $\lambda$ is the freespace wavelength, $\omega_{0}$ is the frequency of the photon mode. $\gamma/2\pi\approx 6.1\,\mathrm{MHz}$ is the spontaneous emission rate of the atom in vacuum, i.e. equal to the angular frequency of natural linewidth. $V_{m}$ denotes the effective mode volume and could be derived approximately $V_{m}\approx2\pi R A_{m}$ for the microring structure with a radius of $R$, and the $A_m$ is the effective mode area. For an atom located at working point $(r_0,z_0)$,
\begin{equation}
A_{m}(r,z)=\frac{\int\varepsilon(r,z)\left|E(r,z)\right|^{2}drdz}{\varepsilon(r_0,z_0)\left|E(r_0,z_0)\right|^{2}},
\end{equation}
where $E(r,z)$ is the electric field distribution and $\varepsilon$ is the distribution of dielectric permittivity.

Therefore, the merit for quantum coherence of the microcavity-enhanced atom-photon interaction can be figured by the cooperativity parameter as
\begin{equation}
    C=\frac{g^2}{\kappa\gamma}=\frac{3\lambda^{3}}{4\pi^{2}}\frac{Q}{V_{m}},
\end{equation}
where $\lambda$ is the wavelength of light in free space. The cooperativity parameter mainly depends on the ratio of quality factor $Q$ and effective mode volume $V_{m}$. High cooperativity is pursued for the high-fidelity quantum gate between atoms and photons, and also potential nonlinear optical effects at single-photon levels. Also, the enhanced photon-atom interaction could also be applied to the single-photon sources, with the cavity enhanced collection efficiency of the radiation from the atomic quantum emitter
\begin{equation}
   \eta=\frac{C}{1+C}.
\end{equation}
Note that $C$ is equivalent to the Purcell factor, and the estimation of the $\eta$ is only valid in the bad-cavity limit with $g<\kappa$~\cite{Berman1994}. If $g$ is too strong, we could increase the cavity linewidth by introducing a higher external coupling rate to the microring cavity to meet the bad-cavity limit.

As studied in the above section, the two-color optical dipole trap could be constructed, and it is feasible to trap single atoms on top of the waveguide around $T( r=R, z=100\,\mathrm{nm})$. We are targeting at the strong atom-photon interaction, thus we are focusing on the fundamental mode polarized along $z$ direction at the wavelength of D2 transitions of Rubidium atom ($\sim780\,\mathrm{nm}$). For the GaN microring with a radius of a few tens of micrometer, the quality factor $Q$ is mainly decided by the material absorption and scattering loss, and $Q=2\times10^6$ with the corresponding $\kappa/2\pi\approx100\,\mathrm{MHz}$ is taken in the simulations~\cite{zheng2020integrated,Stassen:19}.

Figure~\ref{fig5} summarizes the performance of the microcavity enhanced atom-photon interaction in the trap center T($r=R$, $z-h=100\,\mathrm{nm}$) above the microring with different geometry parameters $w$ and $h$. The effective mode area $A_{m}$ and the photon-atom coupling strength $g$ for the $^{87}$Rb and the microring with $R=10\,\mathrm{\mu m}$ and $h=280\,\mathrm{nm}$ are given in Fig.~\ref{fig5}(a). The effective mode area $A_{m}$ increases with increased waveguide width, since a wide waveguide constraints more mode field inside the waveguide with less mode field outside. With an increased effective mode area $A_{m}$, the photon-atom coupling strength $g$ decreases. The cooperativity parameter $C$ of the $^{87}$Rb atom also decrease with increased waveguide width $w$ of microring as shown in Fig.~\ref{fig5} (b). The photon-atom interaction is also related to the waveguide height as shown in Fig.~\ref{fig5}(c) and (d) for the microring with $R=10\,\mathrm{\mu m}$ and $w=600\,\mathrm{nm}$. The effective mode area $A_{m}$ increases and the photon-atom coupling strength $g$ decreases as the waveguide height $h$ increases, since thicker waveguide has less evanescent field outside. The cooperativity parameter C of the
$^{87}$Rb atom also decreases as the waveguide height $h$ increases.

\begin{figure}
\centering \includegraphics[width=1\columnwidth]{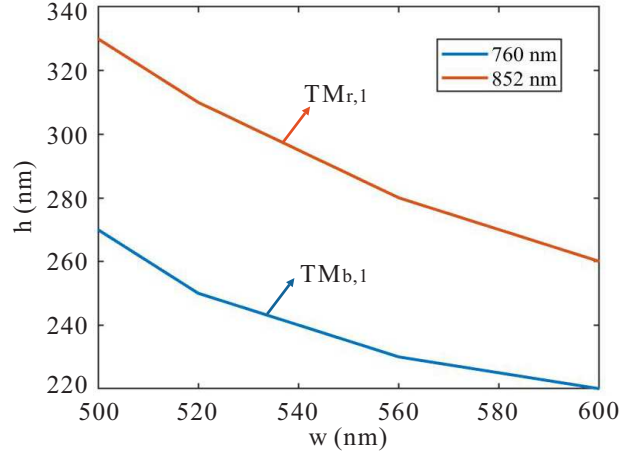}
\caption{The cut-off condition of the $\mathrm{TM_1}$ mode for the blue- and red-detuned wavelength. The lines indicate the crictical parameters for the cut-off condition, and the mode is allowed only when $h$ and $w$ parameters are above the critical line.}
\label{fig6}
\end{figure}

 From the above analysis, the photon-atom interaction increases as the size of microring waveguide decreases. In the same time, the mode loss of the microring cavity will increase as the size of the waveguide decreases. So, an appropriate size of the microring waveguide is important to guarantee the efficient photon-atom interaction. Table~\ref{Table1} gives the photon-atom interaction in the optical trap center T($r=R$, $z-h=100\,\mathrm{nm}$) above the GaN microring waveguide with different parameters incident with $P_{\mathrm{b}}=15\,\mathrm{mW}$ and $P_{\mathrm{b1}}=1.8\,\mathrm{mW}$. It is shown that the depth of optical trap well is larger than 0.13 $\mathrm{mk}$, which can trap single atom stably. In the same time, the trapped atom has an efficient photon-atom interaction with microring waveguide modes. The photon-atom coupling strength $g/2\pi$ is larger than $110\,\mathrm{MHz}$ and the cooperativity parameter $C$ is larger than 42 with the corresponding collection efficiency above $97.7\,\%$. Since the $\mathrm{TM_{b,1}}$ mode is used to improve the trap of the atom, the cut-off of the $\mathrm{TM_1}$ mode is provided in Fig.~\ref{fig6}. For small waveguide width $w$ and height $h$, the ${\mathrm{TM_1}}$ mode can be excited and propagate well in the microring waveguide. So the proposed platform based on suspension-free GaN-on-sapphire chip is feasible to provide stable atom trapping with hybrid modes for the photon-atom interaction in the integrated quantum information.

\begin{table}
\caption{The photon-atom interaction of $^{87}$Rb on the trap well formed above the GaN microring waveguide incident with intracavity trap laser power $P_{\mathrm{b}}=15\,\mathrm{mW}$, $P_{\mathrm{b1}}=1.8\,\mathrm{mW}$, microring radius $R=10\,\mathrm{\mu m}$.}

\begin{tabular}{|c|c|c|c|c|c|}
\hline
$w\,(\mathrm{\mu m})$ & 0.7 & 0.6 & 0.6 & 0.6 & 0.6\tabularnewline
\hline
$h\,(\mathrm{\mu m})$ & 0.28 & 0.32 & 0.3 & 0.28 & 0.28\tabularnewline
\hline
$P_{\mathrm{r}}/P_{\mathrm{b}}$ & 1.27 & 1.25 & 1.25 & 1.32 & 1.36\tabularnewline
\hline
$\begin{array}{c}
T(r-R,z-h)\\
(\mathrm{nm},\mathrm{nm})
\end{array}$ & (1,107) & (-4,103) & (-28,97) & (8,102) & (3,102)\tabularnewline
\hline
$U_{\mathrm{min}}\,(\mathrm{mK})$ & 0.265 & 0.205 & 0.237 & 0.336 & 0.395\tabularnewline
\hline
$U_{\mathrm{eff}}\,(\mathrm{mK})$ & 0.197 & 0.131 & 0.141 & 0.249 & 0.264\tabularnewline
\hline
$A_{\mathrm{m}}\,(\mathrm{\mu m}^{2})$ & 23.5 & 27.4 & 20.0 & 17.1 & 14.7\tabularnewline
\hline
$g/2\pi$ (MHz)) & 118.7 & 110.0 & 128.6 & 139.2 & 150.1\tabularnewline
\hline
$C$ & 48.9 & 42 & 57.4 & 67.2 & 78.2\tabularnewline
\hline
$\eta\,(\%)$ & 98.0 & 97.7 & 98.3 & 98.5 & 98.7\tabularnewline
\hline
\end{tabular}
\label{Table1}
\end{table}

\section{Conclusion}

A stable platform for trapping atoms is proposed to realize the hybrid photon-atom circuit based on a suspension-free GaN-on-sapphire chip. The high-order waveguide modes are used for realizing a stable trap potential well with low incident trap light power, with which a feasible and regular dipole trap potential well around $300\,\mathrm{\mu K}$ can be obtained by only $10\,\mathrm{mW}$-level power circulating in the microring (equivalently sub-mW-level input power in the coupling waveguide). In addition, the interaction between trapped atom and guided photon in the resonator is analyzed to confirm a high-cooperative photon-atom interface for potential applications. Such a stable on-chip atom trapping scheme will play a key role in the hybrid photon-atom circuit, which provides a unique experimental platform for studying the quantum optics effects with novel photonic structure, the single-photon level nonlinear photonics, and the quantum devices for scalable quantum information processing~\cite{chang2018colloquium}.

\section{Acknowledgments}

This work was supported by the National Key Research and Development Program of China (Grant Nos. 2018YFA0306400, 2017YFA0304100), the National Natural Science Foundation of China (Grant Nos. 11922411, U21A6006, 12104441, 12134014, 12074194), the Project funded by China Postdoctoral Science Foundation (SBH190004), the Leading-edge technology Program of Jiangsu Natural Science Foundation (Grant No. BK20192001), and the Fund for Shanxi ``1331 Project'' Key Subjects Construction. C.-L.Z. was also supported by the Fundamental Research Funds for the Central Universities, and the Program of State Key Laboratory of Quantum Optics and Quantum Optics Devices. This work was partially carried out at the USTC Center for Micro and Nanoscale Research and Fabrication.

\end{document}